\documentclass[11pt,a4paper,superscriptaddress,groupedaddress,nofootnoteinbib]{article} 
\usepackage{jheppub}

\usepackage[utf8]{inputenc}
\usepackage{float}
\usepackage{epsfig}
\usepackage{graphicx}% Include figure files
\usepackage{color}
\usepackage{tabularx}
\usepackage{color}
\usepackage{verbatim}
\usepackage[inline,final]{showlabels}
\usepackage{tensor}
\usepackage{enumerate}

\usepackage{subcaption}
\usepackage{color}
\begin{document}
\preprint{TUM-EFT 146/21}
\title{LHC constraints on hidden gravitons}
\author[a]{J.\,A.\,R.\,Cembranos\footnote{cembra@ucm.es}}
\affiliation[a]{Departamento de  Física Teórica and Instituto de Física de Partículas y del Cosmos IPARCOS, Universidad Complutense de Madrid, E-28040 Madrid, Spain}
\author[b]{R.\,L.\,Delgado\footnote{rafael.delgado@upm.es}}
\affiliation[b]{Departamento de Matem\'atica Aplicada a las TIC, ETSIST, Universidad Politécnica de Madrid, E-28040 Madrid, Spain;
Technische Universität München, Physik-Department T30f, James-Franck-Str. 1, 85748 Garching, Germany;
INFN-Firenze, via G. Sansone, 1, 50019 Sesto Fiorentino (FI), Italia}
\author[c]{H. Villarrubia-Rojo\footnote{herojo@phys.ethz.ch}}
\affiliation[c]{Institute for Theoretical Physics, ETH Z\"{u}rich, Wolfgang-Pauli-Strasse 27, 8093, Z\"{u}rich, Switzerland}
\date{\today}

%\begin{abstract}
\abstract{We analyze LHC data in order to constrain the parameter space of new spin-2 particles universally
coupled to the energy-momentum tensor. These new hypothetical particles are the so-called hidden 
gravitons, whose phenomenology at low energies is determined by two parameters: its mass and its 
dimensional coupling constant. Hidden gravitons arise in many different extensions of the 
Standard Model of particles and interactions and General Relativity. Their phenomenology has
been studied mainly in relation to modifications of gravity and astrophysical signatures. In this work,
we extend the constraints for heavy hidden gravitons, with masses larger than $1$~GeV, by 
taking into account events collected by ATLAS and CMS in the WW channel, Drell-Yan processes, 
and the diphoton channel from proton–proton collisions at $\sqrt{s}=8$~TeV.}
%\end{abstract}

\maketitle
%\tableofcontents

\section{Introduction}
	The beginning of the 21st century has witnessed the consolidation of
	two standard models in fundamental physics: the Standard Model (SM) of particle
	physics and the $\Lambda$CDM model of cosmology. Both models have withstood many
	tests over the years and are supported by a large number of extremely precise 
	measurements and observations. Taken together, these two leading models constitute
	a baseline of our understanding of the Universe. However, while their success cannot
	be contested, many questions still remain.
	
	Most of the questions that have prompted the search for extensions to the standard
	paradigm are theoretical in nature: $CP$ problem, origin of neutrino masses, fundamental
	nature of dark matter and dark energy,... Also, as the precision of experiments and 
	observations increase, new puzzles may arise from the observational side, as can
	be illustrated by the growing concern in the cosmological community over the so-called
	$H_0$ tension~\cite{Bernal:2016gxb, Verde:2019ivm}. 
	
	Traditionally, the interplay between particle physics and cosmology has proven to be
	extremely fruitful. Many beyond-SM (BSM) models can be tested based on their cosmological
	implications, e.g. the QCD axion can simultaneously solve the $CP$ problem~\cite{Peccei:1977hh,
	Wilczek:1977pj, Weinberg:1977ma} and act as dark matter~\cite{Sikivie:2009qn, Marsh:2015xka}.
	Similarly, new advances in cosmology can also shed new light on
	particle physics, e.g. large-scale structure surveys expect to measure, at least, the
	sum of the neutrino masses. Hence, the observational implications of any beyond-SM
	or beyond-$\Lambda$CDM model should be carefully analyzed in both realms. 
	
	From the QCD axion to modified gravity theories like
	Horndeski~\cite{Horndeski:1974wa, Deffayet:2009wt, Heisenberg:2018vsk} and 
	Generalized Proca~\cite{Heisenberg:2014rta, Jimenez:2016upj, Heisenberg:2018vsk}, the 
	vast majority of extensions to the standard paradigm
	rely on the inclusion of additional scalar or vector fields, i.e. new spin 0 and 1
	particles. Tensor fields, i.e. spin-2 particles, on the other hand, are commonly
	overlooked. This choice seems reasonable on the grounds of simplicity but not on
	the grounds of naturalness. Since we believe that fundamental spin-2 particles also
	exist in Nature and mediate gravitational interactions, i.e. gravitons, we must
	also explore extensions based on spin-2 fields.
	
	While comparatively less studied in the literature, new massive spin-2 degrees of
	freedom have been shown to arise in different modifications of gravity. 
	Extradimensional theories of gravity, like the ADD~\cite{ArkaniHamed:1998rs,
	Antoniadis:1998ig, ArkaniHamed:1998nn} and Randall-Sundrum~\cite{Randall:1999ee,
	Randall:1999vf, Davoudiasl:1999jd, Garriga:1999yh} 
	models, generically predict the existence of new massive spin-2 particles, either with
	a continuum mass spectrum or as a number of widely separated mass resonances. Also
	in the context of bimetric theories of gravity~\cite{Hassan:2011vm, Schmidt-May:2015vnx,
	Garcia-Garcia:2016dcw} a new massive spin-2 degree of freedom naturally appears. 
	
	The possible existence of new massive gravitons, that we will generically refer to
	as hidden gravitons, prompts the question: what would their observational signature 
	be? This question was partially answered in~\cite{Cembranos:2017vgi}, where different constraints 
	on the mass and coupling of the hidden gravitons were derived, based on their effects
	on fifth-force tests and on stellar energy-loss arguments. In this work we will
	extend these findings to higher masses, where the astrophysical probes are not
	competitive and the signatures in particle colliders set the most restrictive bounds.
	Similar searches were performed in~\cite{Giudice:1998ck, deAquino:2011ix, 
	Tang:2012pv} for specific models and with less updated data.
	
	This work is structured as follows. In section~\ref{sec:theory} we present the 
	details on the theoretical model and the implementation. Section~\ref{sec:LHC}
	contains a description of the different experimental channels and the 
	constraints on the model. Finally, in section~\ref{sec:summary} we summarize
	the main conclusions of the work and show the combined experimental bounds
	on the hidden gravitons.
	
\section{Theoretical framework}\label{sec:theory}
	We will use a generic framework to describe the massive graviton
	\begin{equation}\label{eq:Hid_Mass_FierzPauliL}
        \mathcal{L}_h \equiv-\frac{1}{2}\partial^{\alpha}h^{\mu\nu}(\partial_\alpha h_{\mu\nu}-2 \partial_{(\mu}h_{\nu )\alpha}
                - \partial_\alpha h \eta_{\mu\nu}+2 \partial_{(\mu}h\eta_{\nu )\alpha})-\frac{1}{2}m^{2}(h_{\mu\nu}^{2}-h^{2})\ ,
    \end{equation}
    where $m$ is the graviton mass and $\eta_{\mu\nu}$ is the Minkowski metric.
    This Lagrangian is the well-known Fierz-Pauli Lagrangian~\cite{Fierz:1939ix} that
    describes a massive spin-2 particle. The kinetic and mass terms in this Lagrangian
    can be found imposing the absence of ghost instabilities~\cite{deRham:2014zqa, 
    Heisenberg:2018vsk}. This is the general linear description of a massive spin-2 
    particle, so it can be used as a generic framework to study theories with 
    spin-2 degrees of freedom. We will also choose a universal coupling to the 
    SM for the hidden gravitons, like the standard massless gravitons, 
    \begin{equation}\label{eq:Hid_Mass_interactingL}
    	\mathcal{L} = \mathcal{L}_\text{\tiny SM} + \mathcal{L}_h 
    		+ \kappa h_{\mu\nu}T^{\mu\nu}_\text{\tiny SM}\ ,
    \end{equation}
    where $\kappa$ is the universal coupling to the Standard Model, that can be also 
    rewritten as $\kappa=1/M_h=\sqrt{8\pi G_h}$.
    
    The relevant parton level amplitudes for the subprocesses are $q\bar q\to gG$, $qg\to qG$ and $gg\to gG$, where
    the letters $q$ and $g$ refer generally to quarks and gluons, whereas the $G$ letter stands for the hidden
    graviton. The corresponding cross sections have been studied in different contexts. For example, they can be 
    found on~\cite{Giudice:1998ck}, but for the shake of clarity we reproduce them here:
    \begin{align}
    \frac{d\sigma(q\bar q\to gG)}{dt} &=
    \frac{\alpha_s\kappa^2}{36s}F_1(t/s, m^2/s)\ , \\
    \frac{d\sigma(qg\to qG)}{dt} &=
    \frac{\alpha_s\kappa^2}{96s}F_2(t/s, m^2/s)\ ,\\
    \frac{d\sigma(gg\to gG)}{dt} &=
    \frac{3\alpha_s\kappa^2}{16s}F_3(t/s, m^2/s)\ ;
    \end{align}
    where $s$, $t$ and $u$ are the usual Mandelstam variables for a $2\to 2$ scattering process. The functions
    $F_1$, $F_2$ and $F_3$ are defined by
    \begin{align}
        x(y-1-x)F_1(x,y)    &= (1+4x)y^3 - 6x(1+2x)y^2 + (1+6x+18x^2+16x^3)y  \nonumber  \\
                            &\quad- 4x(1+x)(1+2x+2x^2)\ ,  \\
        x(y-1-x)F_2(x,y)    &= -2y^4 + 4(1+x)y^3 - 3(1+4x+x^2)y^2  \nonumber  \\
                            &\quad+ (1+x)(1+8x+x^2)y -4x(1+x^2)\ , \\
        x(y-1-x)F_3(x,y)    &= y^4 - 2(1+x)y^3+3(1+x^2)y^2-2(1+x^3)y  \nonumber \\
                            &\quad+1+2x+3x^2+2x^3+x^4 \ .
    \end{align}
    
\section{Data analysis}\label{sec:LHC}
    For the computation of the LHC constratins, Pythia~8~\cite{Sjostrand:2006za,Sjostrand:2007gs},
    DELPHES~\cite{deFavereau:2013fsa} and RIVET~\cite{Buckley:2010ar,Bierlich:2019rhm} are used.
    We rely on several \emph{validated} RIVET~\cite{Buckley:2010ar,Bierlich:2019rhm} analysis for comparison with experimental data.
    The BSM processes $gg\to gG$, $qg\to qG$ and $q\bar q\to gG$ are implemented as parton level processes
    inside Pythia~8 framework. We take advantage of the fact that Pythia~8 is implemented as an object oriented C++ library.
    Hence, the new processes are implemented by inheritance of the \verb+Sigma2Process+ class, without modifying the library.
	
	In order to constrain the parameters of the model, we study three observational 
	channels: $H\to WW$ \cite{Khachatryan:2016vnn}, Drell-Yan \cite{Aad:2016zzw} and
	diphoton \cite{Aaboud:2017vol}. Each of them is detailed below. Given that all the
	previous data is compatible with the SM, we will just try to ascertain whether
	the remaining uncertainty in the observations leave room for the simulated signal. 
	We perform a $\chi^2$ test, assuming that the SM background can approximately
	account for the observed data, and then estimating the error as 
	a quadratic sum of the uncertainties in the data and the signal.
	For the computations, we used 175.000h of computer time granted on C2PAP 
	supercomputing facility at the Leibniz Supercomputing Center. Each point in 
	the $m$ vs. $\kappa m$ plots represents 200.000 events generated on Pythia. 
	These events were further processed on Rivet.
	
	\begin{figure}[ht]
		\centering
		\begin{subfigure}[t]{0.48\textwidth}
			\includegraphics[scale=0.55]{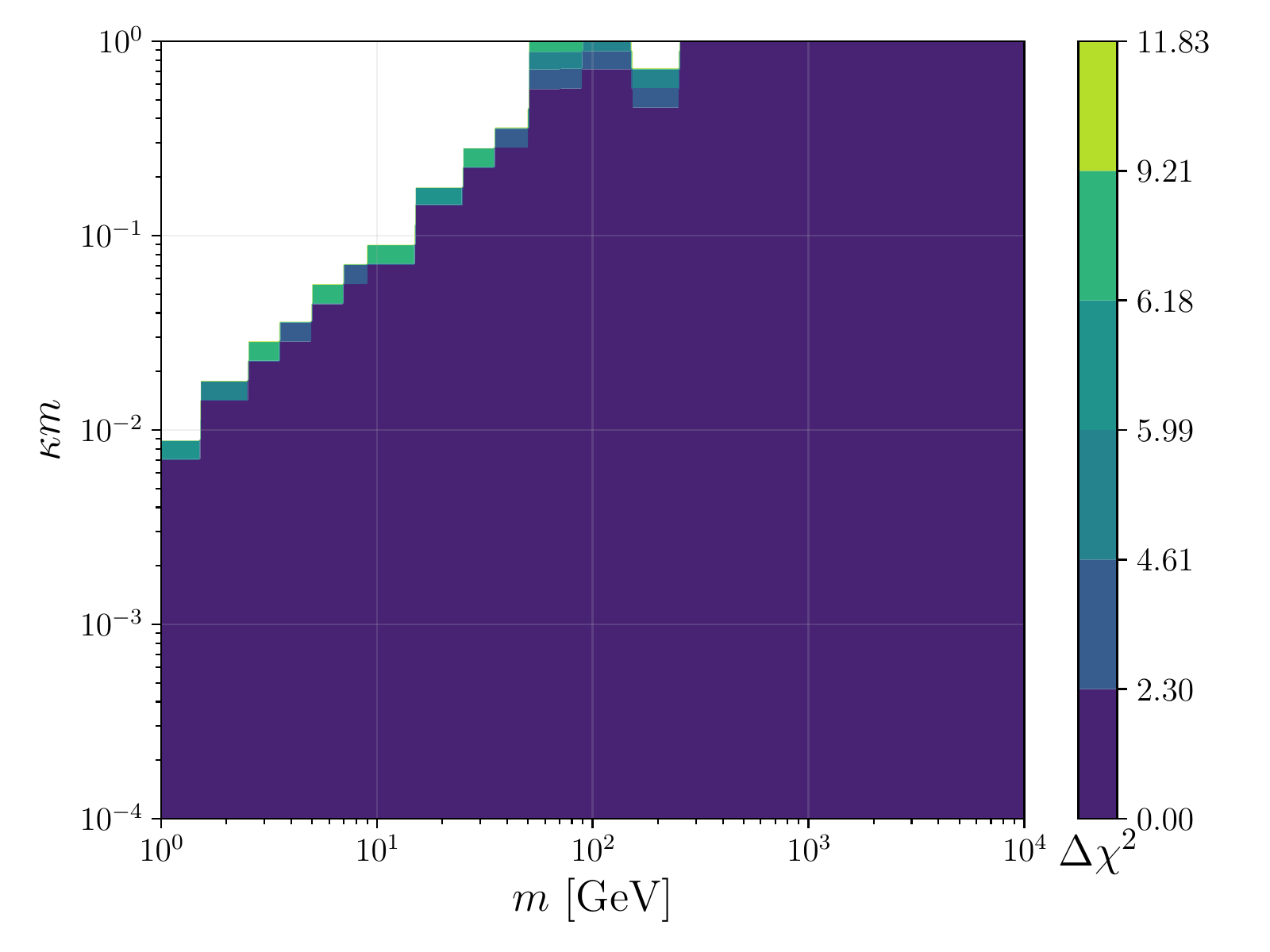}
			\caption{H$\to$WW}\label{fig:HtoWW}
		\end{subfigure}
		\begin{subfigure}[t]{0.48\textwidth}
			\includegraphics[scale=0.55]{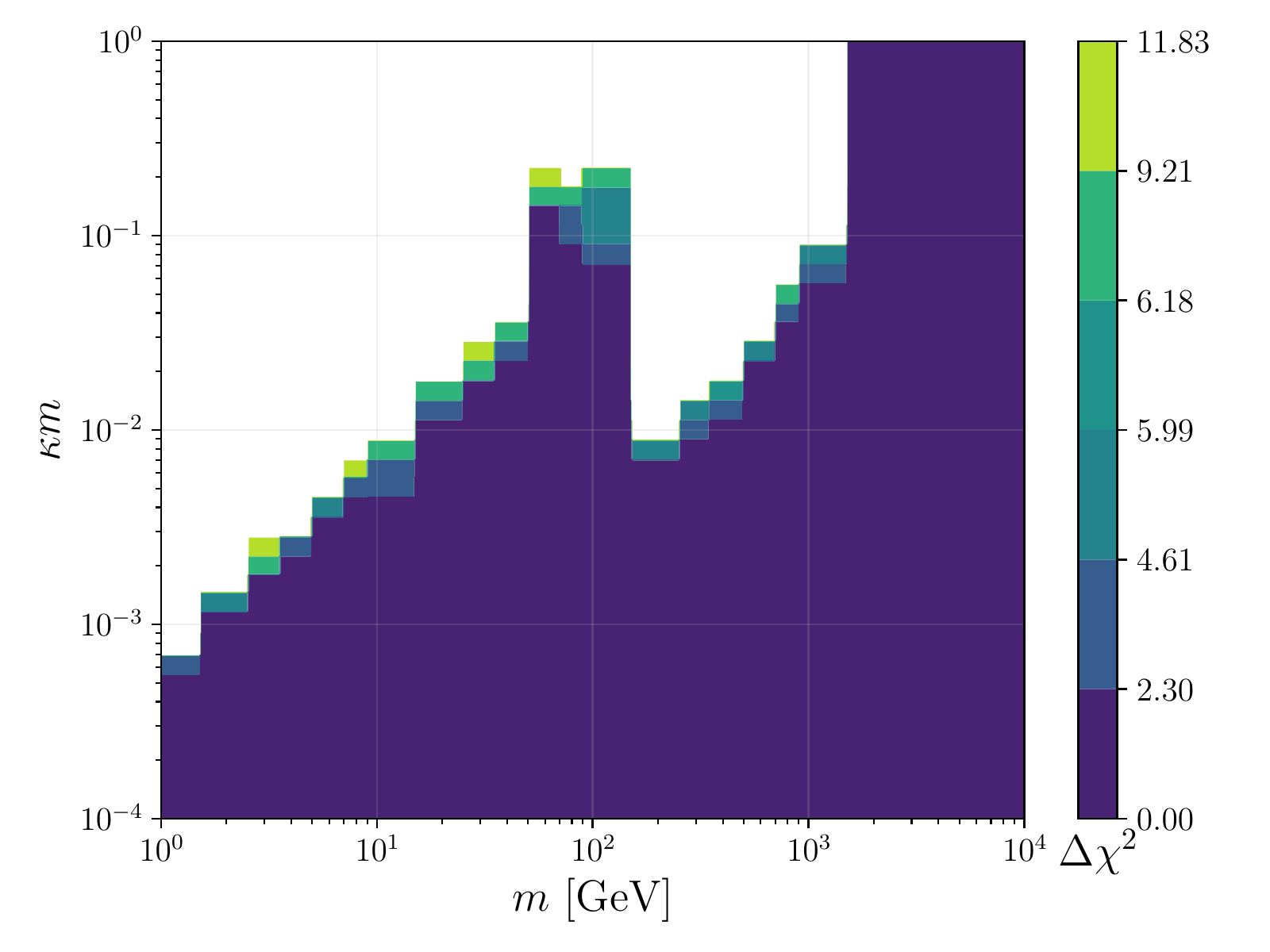}
			\caption{Drell-Yan}\label{fig:DrellYan}
		\end{subfigure}
		\caption{Density plots for the $H\to WW$ process in CMS (left) and Drell-Yan in ATLAS (right). The
			white region is excluded.}
		\label{fig:density_1}
	\end{figure}
	
	\begin{figure}[ht]
		\centering
		\begin{subfigure}[t]{0.48\textwidth}
			\includegraphics[scale=0.55]{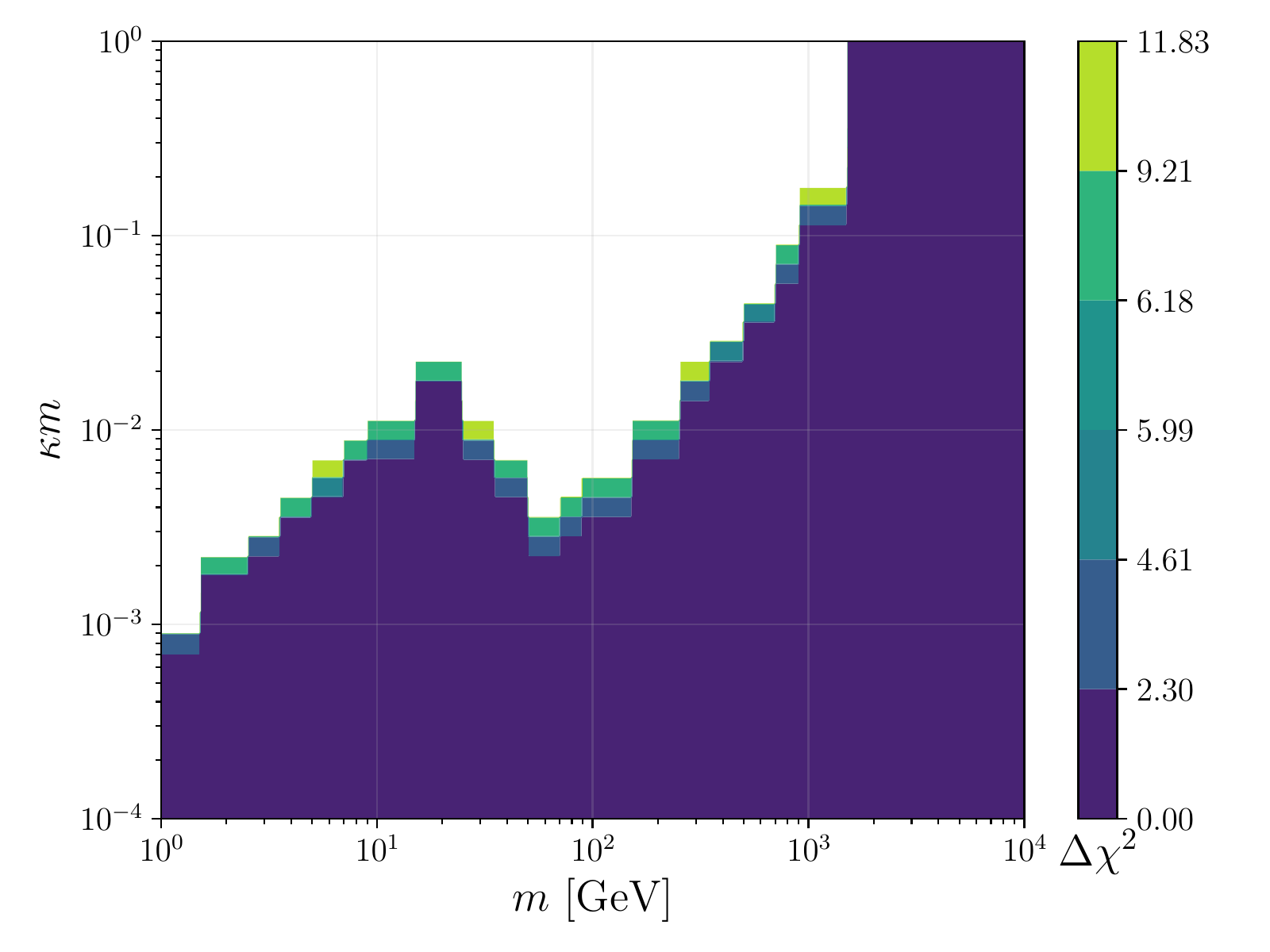}
			\caption{Diphoton (ATLAS)}\label{fig:diphoton}
		\end{subfigure}
		\begin{subfigure}[t]{0.48\textwidth}
			\includegraphics[scale=0.55]{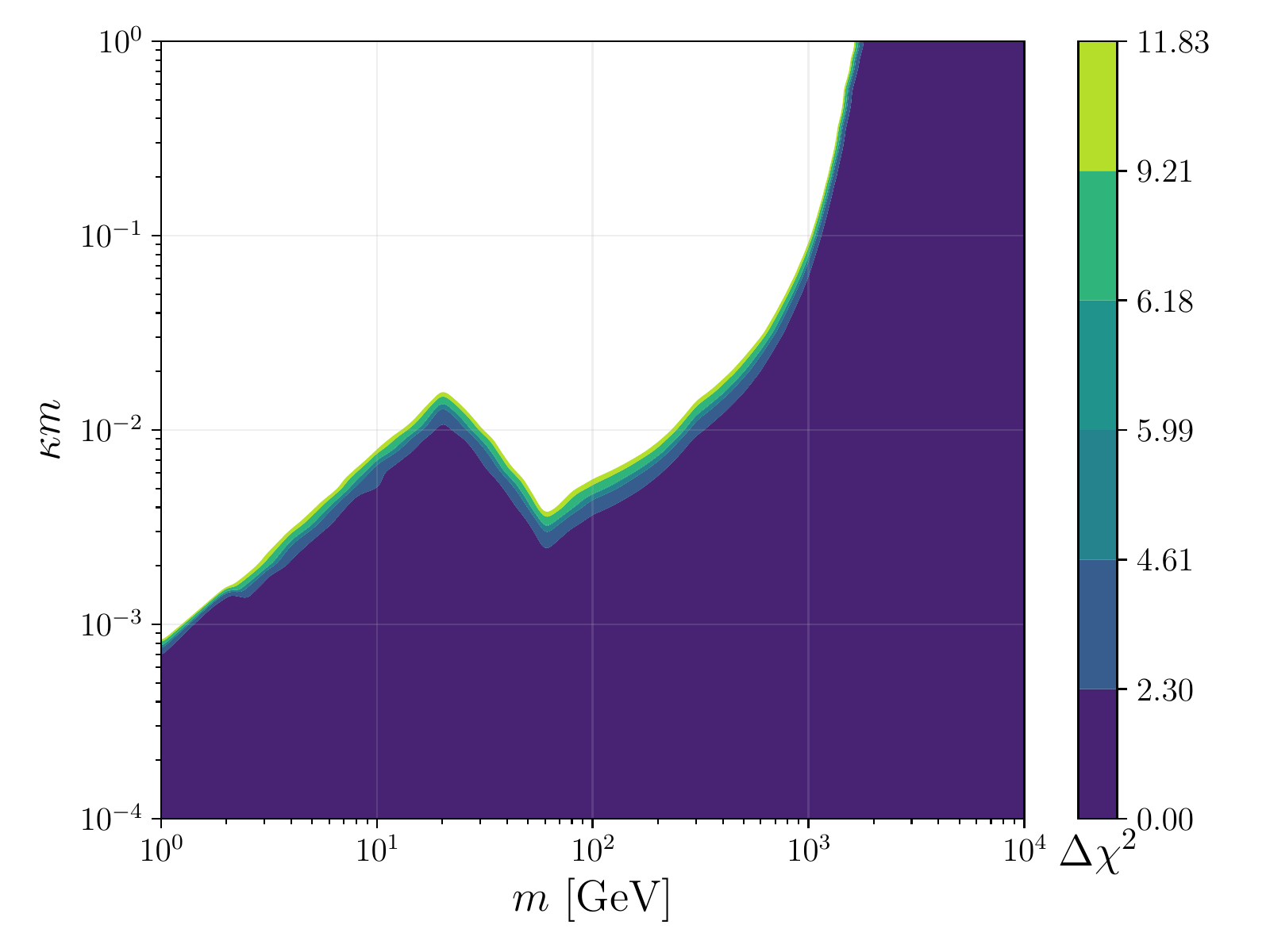}
			\caption{Combined}\label{fig:total}
		\end{subfigure}
		\caption{Density plots for the diphoton production in ATLAS (left) and combined density plot for the
			three channels (right).
			The white region is excluded.}
		\label{fig:density_2}
	\end{figure}

	The first process that we study is the decay \textit{H} to \textit{WW},
	using the CMS\_2017\_I1467451 RIVET analysis (Fig.~\ref{fig:HtoWW}). It models
	the CMS study~\cite{Khachatryan:2016vnn}, based on $H\to WW$ decay channel at $\sqrt{s}=8\,{\rm TeV}$
	(integrated luminosity of $19.4\,{\rm fb}^{-1}$), on the leptonic channel $H\to W^+W^-\to 2l2\nu$.
	The lepton transverse momentum $p_T(ll)$ and missing transverse momentum $p_{T,{\rm miss}}$ are used
	to reconstruct the Higgs transverse momentum $p_T(H)$.	
	The main cuts are: leading lepton, $P_T>20\,{\rm GeV}$; subleading lepton, $P_T>10\,{\rm GeV}$;
	pseudorapidity of electrons and muons, $\lvert\eta\lvert < 2.5$; invariant mass of the two charged leptons,
	$m_{ll}>12\,{\rm GeV}$; charged lepton pair, $p_T>30\,{\rm GeV}$; transverse invariant mass of the leptonic system,
	$m_T^{e\mu\nu\nu}>50\,{\rm GeV}$.
	
	The second channel that we consider in this work is based on 
    the RIVET analysis ATLAS\_2016\_I1467454 (Fig.~\ref{fig:DrellYan}). This is a Drell-Yan study
    in ATLAS~\cite{Aad:2016zzw}, $Z/\gamma^*\to l^+ l^-$ and photo-induced $\gamma\gamma\to l^+ l^-$.
    Integrated luminosity of $20.3\,{\rm fb}^{-1}$ at $\sqrt{s}=8\,{\rm TeV}$. 
    For both the electron and muon channels, the cut over the invariant mass of the lepton pairs
    is $116\,{\rm GeV}<m_{ll}<1500\,{\rm GeV}$.    
    The electron channel has a cut of $E_T(e)>40\,{\rm GeV}$ for the leading electron and $E_T(e)>30\,{\rm GeV}$
    for the subleading one. The pseudorapidities are in the range $\lvert\eta^e\rvert < 2.47$, excluding $1.37<\lvert\eta^e\rvert<1.52$.
    The absolute difference in pseudorapidity between the two electreos is restricted to $\lvert\Delta\eta_{ee}\rvert < 3.5$.
    Concerning the muon channel, at least two oppositely charged muones with transverse momenta
    $p_T^\mu>40\,{\rm GeV}$ (leading muon) and $p_T^\mu>30\,{\rm GeV}$ (subleading muon) are required.
    The pseudorapidity should be $\lvert\eta^\mu\rvert < 2.4$. No requirement is placed on $\Delta\eta_{\mu\mu}$.

	Finally, we also include 
	the RIVET analysis ATLAS\_2017\_I1591327 (Fig.~\ref{fig:diphoton}),  that corresponds to
	the diphoton production in ATLAS~\cite{Aaboud:2017vol} at $\sqrt{s}=8\,{\rm TeV}$ and integrated luminosity
	of $20.2\,{\rm fb}^{-1}$. The cuts are: transverse energies $E_{T,1}^\gamma>40\,{\rm GeV}$ (leading photon)
	and $E_{T,2}^\gamma>30\,{\rm GeV}$ (subleading one) and pseudorapidities $\lvert\eta^\gamma\rvert<1.37$
	or $1.56<\lvert\eta^\gamma\rvert<2.37$.\\
	
	The combined constraints from the three processes can be found on 
	Fig.~\ref{fig:total}. On Fig.~\ref{fig:combined} we compare these new collider
	constraints on hidden gravitons 
	with those of astrophysical and 5th force tests~\cite{Cembranos:2017vgi}.

\section{Summary and conclusions}\label{sec:summary}
    The results of the commented analyses are translated into exclusion limits on the mass and
    the coupling of the hidden graviton. The sensitivity of the constraints are limited by
    the effect of experimental uncertainties related to jet and transverse missing enery scales and resolutions.
    The choice of different PDF sets results in up to $\sim 10\%$ order of magnitude uncertainties in the acceptance
    and in the cross section. Varying the renormalization and factorization scales introduces $\sim 5\%$ variations
    of the cross section and acceptance. In addition, the uncertainty in the integrated luminosity is included.
    Fig.~\ref{fig:density_1} shows the derived $95\%$ CL exclusion limits in the mentioned $\kappa-m$ parameter space
    of the hidden gravitons for the WW channel (left panel) and Drell-Yan process (right panel). The same bounds
    are plotted for the diphoton channel in Fig.~\ref{fig:density_2} (left panel). The combined results from the tree analyses
    are shown in Fig.~\ref{fig:density_2} (right panel). These combined constraints are dominated by the diphoton data.
    
\begin{figure}[h]
    		\includegraphics[scale=0.7]{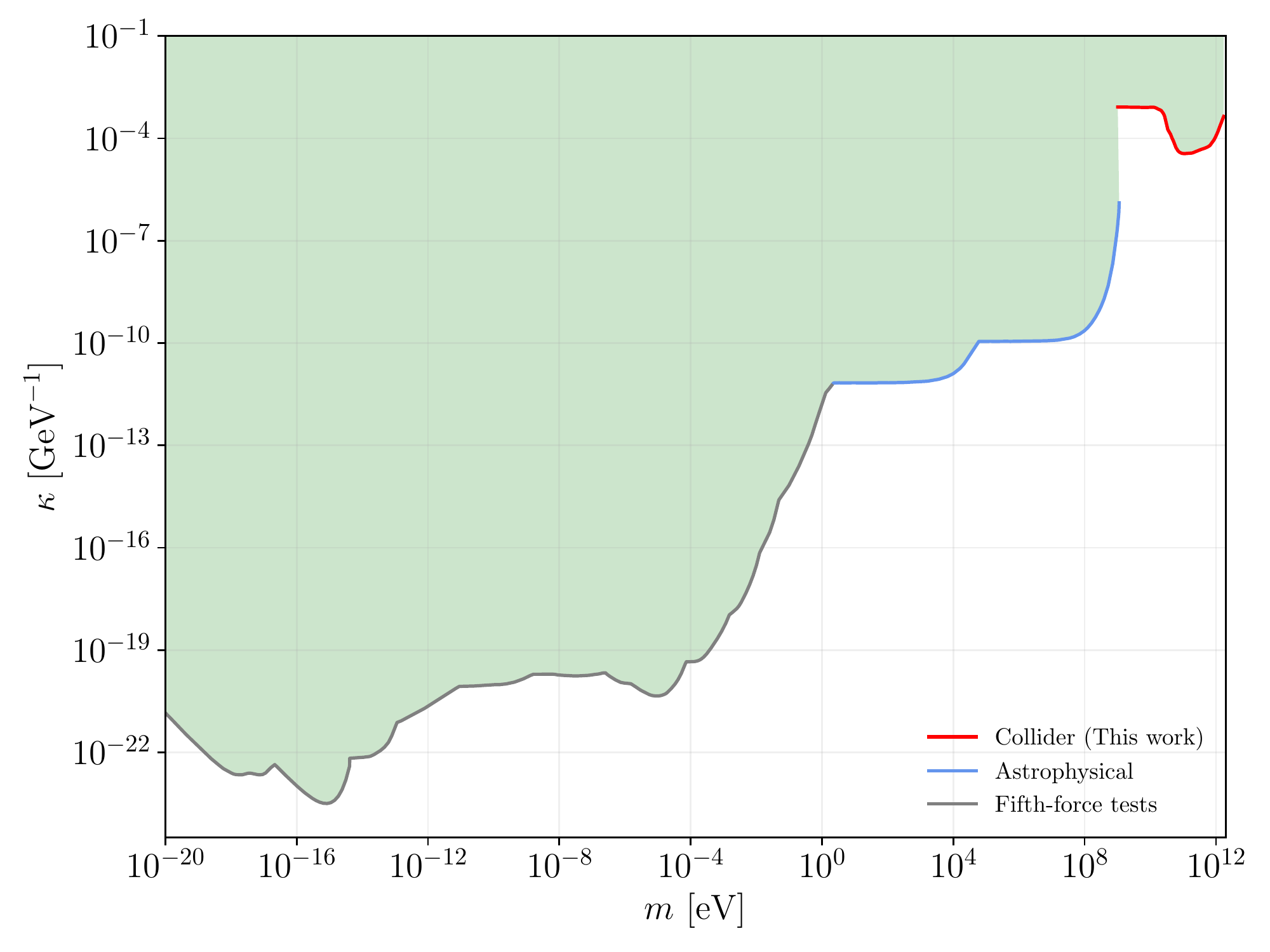}
    		\caption{Total constraints in the hidden-graviton parameter space. The shaded
    		region is excluded. ``Fifth-force tests'' represents a 
    		collection of laboratory and Solar System experiments, see \cite{Cembranos:2017vgi}
    		and references therein. The astrophysical bounds were derived in
    		\cite{Cembranos:2017vgi}, based on stellar energy-loss arguments.  
    		``Collider'' represents the combined bounds from Figures \ref{fig:density_1} and \ref{fig:density_2}.}\label{fig:combined}
    	\end{figure}

    These results are translated into the general paramater space of hidden gravitons presented 
    in Fig.~\ref{fig:combined}, where it is possible to see that they are the most constraining for heavy gravitons,
    i.e. for hidden graviton masses larger than $m\sim 1$~GeV. The phenomenology of gravitons with masses between
    $m\sim 1$~eV and $m\sim 1$~GeV is more limited by astrophysical data~\cite{Cembranos:2017vgi}, whereas
    light hidden gravitons ($m\lesssim 1$~eV) suffer important restrictions from fifth force experiments~\cite{Cembranos:2017vgi}.
    
    In summary, results are reported from a search for hidden gravitons in events associated with 
    the WW channel, Drell-Yan processes, and the diphoton channel from proton–proton collisions at
    $\sqrt{s}=8$~TeV at the LHC, based on data corresponding to an integrated luminosity close to
    $20$~fb$^{-1}$ collected by the ATLAS (Drell-Yan and diphoton) and CMS (WW) experiments. 
    The measurements are in agreement with the SM predictions. The results are translated into
    model-independent 95$\%$ confidence-level limits on the universal hidden graviton coupling
    depending on its mass. The comparison with previous analyses shows that the constraints
    derived in this study are the most important for heavy hidden gravitons.

\begin{acknowledgments}
	We would like to thank Antonio López Maroto for valuable discussions at the beginning
	of this study.
    R.L.D. was financially supported by the Ramón Areces Foundation, the INFN post-doctoral fellowship
    AAOODGF-2019-0000329 and the Spanish grant MICINN: PID2019-108655GB-I00. 
    HVR  is supported by funding from the European Research Council (ERC) under the European Unions Horizon 2020 
    research and innovation programme grant agreement No 801781.    
    The simulations
    have been carried out on the computing facilities of the Computational Center for Particle and Astrophysics (C2PAP)
    and the Leibniz Supercomputing Center (SuperMUC), on the local theory cluster (T30 cluster)
    of the Physics Department of the Technische Universität München (TUM), and on local computing facilities at the INFN-Firenze.
\end{acknowledgments}

\bibliographystyle{JHEP}
\bibliography{Biblio.bib}

\end{document}